\documentclass[letterpaper,twocolumn,english,aps,pra,floatfix,showpacs,tightenlines]{revtex4}
\usepackage[T1]{fontenc}
\usepackage[latin9]{inputenc}
\usepackage{color}
\usepackage{amsmath}
\usepackage{amssymb}
\usepackage{graphicx}
\usepackage{esint}

\makeatletter

\pdfpageheight\paperheight
\pdfpagewidth\paperwidth

\@ifundefined{textcolor}{}
{%
 \definecolor{BLACK}{gray}{0}
 \definecolor{WHITE}{gray}{1}
 \definecolor{RED}{rgb}{1,0,0}
 \definecolor{GREEN}{rgb}{0,1,0}
 \definecolor{BLUE}{rgb}{0,0,1}
 \definecolor{CYAN}{cmyk}{1,0,0,0}
 \definecolor{MAGENTA}{cmyk}{0,1,0,0}
 \definecolor{YELLOW}{cmyk}{0,0,1,0}
 }

\usepackage{hyperref}

\usepackage{times}

\newcommand{\tr}{\mathrm{tr}}

\newcommand{\1}{\leavevmode{\rm 1\ifmmode\mkern  -4.8mu\else\kern -.3em\fi I}}

\usepackage{babel}

\makeatother

\usepackage{babel}
\begin{document}

\title{Equilibration times in clean and noisy systems}

\author{Lorenzo Campos Venuti, Sunil Yeshwanth and Stephan Haas}

\affiliation{Department of Physics and Astronomy and Center for Quantum Information
Science \& Technology, University of Southern California, Los Angeles,
California 90089-0484, USA}
\begin{abstract}
We study the equilibration dynamics of closed finite quantum systems
and address the question of the time needed for the system to equilibrate.
In particular we focus on the scaling of the equilibration time $T_{\mathrm{eq}}$
with the system size $L$. For clean systems we give general arguments
predicting $T_{\mathrm{eq}}=O\left(L^{0}\right)$ for clustering initial
states, while for small quenches around a critical point we find $T_{\mathrm{eq}}=O\left(L^{\zeta}\right)$
where $\zeta$ is the dynamical critical exponent. We then analyze
noisy systems where exponentially large time scales are known to exist.
Specifically we consider the tight-binding model with diagonal impurities
and give numerical evidence that in this case $T_{\mathrm{eq}}\sim Be^{CL^{\psi}}$
where $B,C,\,\psi$ are observable dependent constants. Finally, we
consider another noisy system whose evolution dynamics is randomly
sampled from a circular unitary ensemble. Here, we are able to prove
analytically that $T_{\mathrm{eq}}=O\left(1\right)$, thus showing
that noise alone is not sufficient for slow equilibration dynamics.
\end{abstract}

\pacs{03.65.Yz, 05.30.-d}

\maketitle

\section{Introduction}

Consider a quantum system initialized in a given state and then allowed
to evolve undisturbed under the action of a time independent Hamiltonian.
Experimentally accessible quantities are expectation values of physical
observables $A$ at a given time $\langle A(t)\rangle$. The timescale
at which such an expectation value relaxes to equilibrium identifies
the \emph{equilibration time }of the particular dynamics. In infinite
systems, equilibration times are typically extracted from the exponential
decay of observables or correlation functions towards their equilibrium
values. For finite systems, or more generally for systems with discrete
energy spectrum, the dynamics is quasi-periodic and such exponential
decay cannot occur, thus necessitating a different definition of equilibration
time. In these cases a meaningful definition of equilibration time
is the first time instant for which the value of an observable equals
its equilibrium value, i.e.~given an observable $A$ and its equilibrium
value $\bar{A}$, $T_{\mathrm{eq}}$ is the smallest $t$ for which
$A(t)=\bar{A}$ \cite{brandino_quench_2012}.

In this paper we study the behavior of equilibration times according
to this definition for various physical systems. In particular we
are interested in the scaling behavior of the equilibration time as
a function of the linear size of the system %
\footnote{Of course proper units of equilibration times are set by $\hbar/J$
where $J$ is an energy-scale of the system.%
}. 

We first consider equilibration times in clean systems. Using the
Loschmidt echo as a particular observable, we are able to give general
estimates for the scaling of the $T_{\mathrm{eq}}$ as a function
of length. Given a local Hamiltonian $H=\sum_{i}h_{i}$ and a sufficiently
clustering initial state $|\psi_{0}\rangle$, i.e.~$g_{i,j}=\langle h_{i}h_{j}\rangle-\langle h_{i}\rangle\langle h_{j}\rangle$
decays sufficiently fast as a function of $i-j$ at large separations,
we find that $T_{\mathrm{eq}}=O(L^{0})$, i.e.~the equilibration
time is independent of the system size. This scaling becomes $T_{\mathrm{eq}}=O(L^{\zeta})$
where $\zeta$ is the dynamical critical exponent in case both the
initial state and the evolution Hamiltonian are close to a critical
point. These general findings are checked explicitly for the Ising
chain in a transverse field. However, one should be cautious that
these arguments may need modifications for other observables e.g.~such
as those undergoing spontaneous symmetry breaking. 

We then turn our attention to systems with random impurities. According
to intuition based on classical models, one expects, in general, a
slower transient approach to equilibrium in noisy systems compared
to clean systems. Indeed, it is known that exponentially large time-scales
are present in glassy systems \textcolor{black}{\cite{mcmillan_scaling_1984}}.
To check and further understand this conjecture, we compute numerically
the equilibration time for the tight-binding model with diagonal impurities,
sometimes called Anderson model \cite{anderson_absence_1958}. \textcolor{black}{A
similar model (albeit with pseudo-random diagonal elements) has recently
been studied in }\cite{gramsch_quenches_2012}\textcolor{black}{,
where it was found that observables relax following a power-law behavior.}
\textcolor{black}{Such power-law equilibration pattern (observed also
in \cite{khatami_quantum_2012}for another noisy system) is itself
a signature of slow equilibration. However, according to our definition,
equilibration times also depend on the long-time equilibrium value.
Our findings confirm a very slow equilibration dynamics characterized
by equilibration times diverging exponentially with the system size. }

\textcolor{black}{Disorder alone might not be sufficient to guarantee
the presence of exponentially large relaxation timescales. To illustrate
this point, we analyze a second noisy system. In this case the evolution
operator is related to a unitary matrix sampled from the circular
unitary ensemble (CUE). }This model is similar to the ones previously
considered in \cite{brandao_convergence_2011,cramer_thermalization_2012},
where, according to a different definition, equilibration times decreasing
algebraically with the size were predicted \cite{cramer_thermalization_2012}.
Using our definition we prove analytically that, for the Loschmidt
echo, $T_{\mathrm{eq}}=O(1)$. 

The paper is organized as follows. In Section II, we describe a clean
system and the observables and the quench considered to study the
nature of equilibration. Our numerical and analytical results for
equilibration timescales and the scaling of these timescales with
system length are outlined for the cases of the tight-binding model
in Section III and the CUE based evolution in Section IV, respectively.
We present our conclusions in Section V.

\section{Equilibration in clean systems}

Before considering noisy systems, let us recall some elementary yet
important facts regarding unitary equilibration in clean systems.
The system is initialized in some state $\rho_{0}$ which evolves
unitarily via $\rho\left(t\right)=e^{-itH}\rho_{0}e^{itH}$. First
note that, because of the unitary nature of the evolution, $\rho\left(t\right)$
does not converge in the strong sense as $t\to\infty$. This is true
irrespective of the Hilbert space dimension, i.e.~also in the thermodynamic
limit. One can then consider the possibility of a weaker convergence
by looking at {}``matrix elements'' $\mathcal{A}\left(t\right)=\tr\left[A\rho\left(t\right)\right]$
where $A$ is an observable. In the thermodynamic limit the spectrum
becomes continuous and one \emph{can} have limit $\mathcal{A}\left(\infty\right)=\lim_{t\to\infty}\mathcal{A}\left(t\right)$
for some observables $A$ (or appropriately rescaled observables)
essentially as a consequence of Riemann-Lebesgue lemma (see e.g.~\cite{campos_venuti_unitary_2010}
and also \cite{ziraldo_relaxation_2012} for a recent discussion).
For finite systems however, $\mathcal{A}\left(t\right)$ is a trigonometric
polynomial and hence, once again, does not admit an infinite time
limit. For the same reason though, the time average $\overline{\mathcal{A}}:=\lim_{T\to\infty}\int_{0}^{T}\mathcal{A}\left(t\right)dt$
exists and is finite. Such a time average coincides with the infinite
time limit when the latter exists, i.e.~$\overline{\mathcal{A}}=\mathcal{A}\left(\infty\right)$.
So the time average can be seen as a mathematical trick, reminiscent
of Cesaro summation, to obtain the infinite time limit when the function
oscillates. Alternatively the time average can mimic the actual measurement
process. In this case $T$ is the {}``observation time'' which one
may argue to be very large compared to the time scales of the unitary
dynamics (see e.g.~\cite{huang_statistical_1963}). Clearly one may
want to investigate the effect of a finite $T$, here for simplicity
we will always take $T\to\infty$. 

\begin{figure}
\begin{centering}
\includegraphics[clip,scale=0.2]{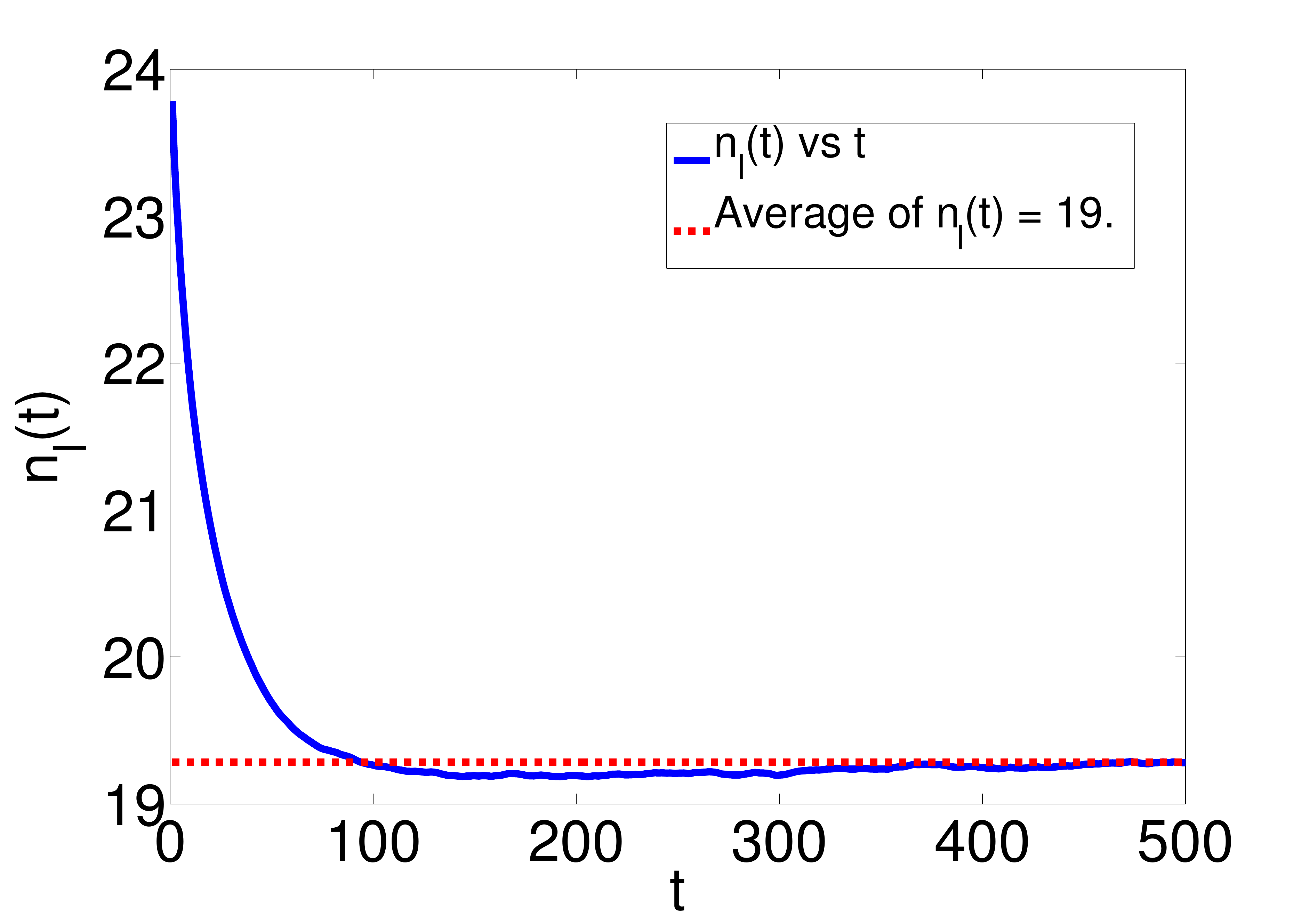}
\par\end{centering}

\caption{Relaxation of $\mathsf{n}_{\ell}\left(t\right)=\mathsf{E}\left[N_{\ell}\left(t\right)/L\right]$
with time for lengths $L=96$, at filling $\nu=N/L=0.25$ and $\alpha=\ell/L=0.5$.
The random distribution has variance $\sigma=0.6$ ($W=0.6\sqrt{3}\approx1.03$).
The ensemble average is computed using 2000 realizations at each time.\label{fig:A_vs_time_osc}}
\end{figure}

Because of the above considerations, the standard definition of \emph{equilibration
time}, extracted from the exponential decay of some time dependent
observable or correlation function, does not make sense for finite
systems, because the dynamics is almost periodic and no exponential
decay is possible. Instead, in finite systems, expectation values
$\mathcal{A}\left(t\right)$ start from a value $\mathcal{A}\left(0\right)$
which retains memory of the initial state $\rho_{0}$, and then, after
an \emph{equilibration time}, approach an average value $\overline{\mathcal{A}}$
and start fluctuating around it with fluctuations $\delta\mathcal{A}=\sqrt{\overline{\mathcal{A}^{2}}-\overline{\mathcal{A}}^{2}}$
due to the finite dimensionality of the system. Clearly the precise
concept of equilibration time is to some extent arbitrary, and different
definitions are possible. Ours is the following: $T_{\mathrm{eq}}$
is the first time for which $\mathcal{A}\left(t\right)$ equals the
average $\overline{\mathcal{A}}$, i.e.~is the first solution of
$\mathcal{A}\left(T_{\mathrm{eq}}\right)=\overline{\mathcal{A}}$
(see, for example, fig.~\ref{fig:A_vs_time_osc}). This definition
is both simple to implement and physically clear. 

Some comments are in order: i) Clearly the precise numerical value
of $T_{\mathrm{eq}}$ is irrelevant, whereas the important information
is contained in the scaling dependence of $T_{\mathrm{eq}}$ on the
system size, $T_{\mathrm{eq}}\left(L\right)$. ii) In principle $\mathcal{A}\left(t\right)$
could intersect $\overline{\mathcal{A}}$ at a first time $T_{1}$,
deviate considerably from the average and intersect $\overline{\mathcal{A}}$
again at $T_{2}$, and possibly have multiple intersections up to
$T_{n}$ before fluctuations of order $\delta\mathcal{A}$ start to
set in. In this situation it seems that the equilibration process
cannot be captured by a single $T_{\mathrm{eq}}$ but rather consists
of many time scales. In all the situations encountered in our analysis,
however, we found that equilibration could always be captured by a
single time-scale. iii) The above definition depends on the specific
observable $A$ and different observables may have, in principle,
different equilibration times. However, we expect that for large classes
of observables the scaling dependence with $L$ will be the same.
iv) Based on the experience with infinite systems, where observables
decay towards their equilibrium value, one could define $T_{\mathrm{eq}}$
as that time for which $\mathcal{A}\left(t\right)$ is \emph{off by
a small} amount from its average value $\overline{\mathcal{A}}$.
This is, after all, the definition we use in the thermodynamic limit
when the approach to equilibrium is exponential. A definition of this
sort has been used for instance in \cite{cramer_thermalization_2012}.
In this way, however, $T_{\mathrm{eq}}$ depends in principle on the
way we define \emph{small. }

To test the equilibration time we consider a particular observable,
the Loschmidt echo (LE) which we define as 
\begin{equation}
\mathcal{L}\left(t\right)=\left|\langle\psi_{0}|e^{-itH}|\psi_{0}\rangle\right|^{2}\,,\label{eq:LE}
\end{equation}
where $|\psi_{0}\rangle$ is the state in which we initialize the
system and $H$ the evolution Hamiltonian. The Loschmidt echo has
been first introduced in the context of quantum chaos (see e.g.~\cite{gorin_dynamics_2006}),
and is generally given a more general form in that context. Equation
(\ref{eq:LE}) can be seen as the time evolved expectation value of
the observable $|\psi_{0}\rangle\langle\psi_{0}|$, and is also known
as \emph{survival probability}. Here we consider the LE because it
is amenable of a cumulant expansion which correctly approximates $\mathcal{L}\left(t\right)$
for sufficiently large times of the order of $T_{\mathrm{eq}}$. This
conclusion is based on numerical experiments on the Ising model in
transverse field \cite{campos_venuti_unitary_2010}. A hand waving
argument is the following: since $\mathcal{L}$ is morally a product
of $L^{d}$ terms (see below) it should be clear that Taylor expansion
of $\ln\mathcal{L}$ works better than the Taylor expansion of $\mathcal{L}$
itself. The cumulant expansion of Eq.~(\ref{eq:LE}) reads
\begin{equation}
\mathcal{L}\left(t\right)=\exp\left[2\sum_{n=1}^{\infty}\frac{\left(-t^{2}\right)^{n}}{\left(2n\right)!}\langle H^{2n}\rangle_{c}\right]\,,\label{eq:LE-cumulant}
\end{equation}
 where $\langle\cdot\rangle_{c}$ stands for connected average with
respect to $|\psi_{0}\rangle$. Truncating Eq.~(\ref{eq:LE-cumulant})
up to the first order we obtain $\mathcal{L}\left(t\right)\simeq\exp\left[-t^{2}\left(\Delta H^{2}\right)\right]$
($\Delta H^{2}=\langle H^{2}\rangle-\langle H\rangle^{2}$). Equating
the short time expansion to the average value $\overline{\mathcal{L}}$
we get the following expression for the equilibration time:
\begin{equation}
T_{\mathrm{eq}}=\sqrt{\frac{-\ln\overline{\mathcal{L}}}{\Delta H^{2}}}\,.\label{eq:T_eq_LE}
\end{equation}

As we will see, the above estimate for $T_{\mathrm{eq}}$ works well
for the Loschmidt echo Eq.~(\ref{eq:LE}). Some comments are in order
at this point. First of all, the equilibration time in Eq.~(\ref{eq:T_eq_LE})
is inversely proportional to the square root of an energy fluctuation.
This is not simply the inverse of an energy gap as one might guess
na\"ively. Secondly, we would like to compare Eq.~(\ref{eq:T_eq_LE}),
with another estimate of equilibration time which appeared recently
in a single body setting \cite{yurovsky_dynamics_2011}. The estimate
of \cite{yurovsky_dynamics_2011} reads $T_{\mathrm{eq}}\sim1/\langle\Delta E_{\mathrm{min}}\rangle_{\mathrm{ave}}$,
where $\Delta E_{\mathrm{min}}$ is the minimum energy gap averaged
over an energy shell around the initial energy $\langle H\rangle$
\footnote{V.~Yurovsky private communication.%
}. So, apart from the order of averages, the equilibration time in
\cite{yurovsky_dynamics_2011} is inversely proportional to a standard
deviation of an energy fluctuation as much as in Eq.~(\ref{eq:T_eq_LE}).
The definitions differ in the numerator which takes into account the
many-body nature of the problem. Thirdly, the estimate Eq.~(\ref{eq:T_eq_LE})
is valid only for the equilibration time of the Loschmidt echo, and
in principle, different observables might equilibrate with different
time scales. 

We will now provide arguments to estimate Eq.~(\ref{eq:T_eq_LE})
which first appeared in \cite{campos_venuti_unitary_2010}. For a
local Hamiltonian $H=\sum_{i}h_{i}$ and a sufficiently clustering
initial state $|\psi_{0}\rangle$, i.e.~$g_{i,j}=\langle h_{i}h_{j}\rangle-\langle h_{i}\rangle\langle h_{j}\rangle$
decays sufficiently fast as a function of $i-j$ at large separations,
all the cumulants in Eq.~(\ref{eq:LE-cumulant}) are extensive in
the system size, that is in a $d$-dimensional system of linear size
$L$, $\langle H^{2n}\rangle_{c}\propto L^{d}$. This means that at
leading order in $L$, $\mathcal{L}\left(t\right)\simeq e^{f\left(t\right)L^{d}}$
\footnote{Alternatively one can use the quantum-classical mapping to show that
at imaginary time, $\mathcal{L}$ is the partition function of a certain
classical system see e.g.~\protect{\cite{gambassi_statistics_2011,campos_venuti_universal_2009}}.%
} and so, by Jensen's inequality $e^{\overline{f}L^{d}}\le\overline{\mathcal{L}}\le\exp[L^{d}\max_{t}f\left(t\right)]$,
showing that $\overline{\mathcal{L}}$ is exponentially small in the
system volume (note that we must have $f\left(t\right)\le0$). For
this reason it is sometimes useful to consider the logarithm of the
LE $\mathcal{F}\left(t\right)=\ln\mathcal{L}\left(t\right)$. The
equilibration time for $\mathcal{F}$ would then be given by $T_{\mathrm{eq}}^{\mathcal{F}}=\sqrt{-\overline{\ln\mathcal{L}}/\Delta H^{2}}$.
Now, since $\overline{\ln\mathcal{L}}\simeq\overline{f\left(t\right)}L^{d}$,
we see that the equilibration times for $\mathcal{L}$ and $\mathcal{F}=\ln\mathcal{L}$
are expected to give the same scaling with respect to $L$. From equation
(\ref{fig:Teq_LE}) we get then $T_{\mathrm{eq}}=O\left(1\right)$:
the equilibration time is independent of the system size. This argument
fails when one considers small quenches close to a quantum critical
point as in this case the clustering properties of the initial state
tends to break down. In this case $|\psi_{0}\rangle$ is the ground
state of $H_{0}=H\left(\lambda_{0}\right)$ where the external parameter
$\lambda_{0}$ is close to a quantum critical point. One then suddenly
changes the parameters by a small amount $\lambda_{0}\to\lambda_{0}+\delta\lambda$
and the system is let evolve undisturbed with Hamiltonian $H=H\left(\lambda_{0}+\delta\lambda\right)$.
In the very small quench regime, roughly $\delta\lambda\ll\min\{L^{-1/\nu},L^{-2/d}\}$
where $\nu$ is the correlation length exponent, perturbation theory
is applicable (see below) and the average LE reduces to $\overline{\mathcal{L}}\approx\left|\langle\psi_{0}|0\rangle\right|^{4}$,
where $|0\rangle$ is the ground state of $H$ \cite{rossini_decoherence_2007}.
The scaling properties of the fidelity $\left|\langle\psi_{0}|0\rangle\right|$
close to a quantum critical point have been studied in \cite{campos_venuti_quantum_2007}
where it was shown that $\left|\langle\psi_{0}|0\rangle\right|\approx1-\mathrm{const.}\times\delta\lambda^{2}L^{2\left(d+\zeta-\Delta\right)}$,
where $\zeta$ is the dynamical critical exponent, and $\Delta$ the
scaling dimension of the operator driving the transition. Using similar
scaling arguments one can show that the variance scales as $\langle H^{2}\rangle_{c}\approx L^{2\left(d-\Delta\right)}$
\cite{campos_venuti_unitary_2010}. These results are valid in the
perturbative regime where $\left|\langle\psi_{0}|0\rangle\right|$
is not too far from 1, i.e.~$\delta\lambda\ll\min\{L^{-1/\nu},L^{-2/d}\}$
where $\nu=\left(d+\zeta-\Delta\right)^{-1}$ \cite{campos_venuti_quantum_2007}.
From equation (\ref{fig:Teq_LE}) we now get $T_{\mathrm{eq}}=O\left(L^{\zeta}\right)$,
i.e.~the equilibration time diverges for large systems. Such a divergence
is reminiscent of the critical slowing down observed in quantum Monte
Carlo algorithms.

\subsection{Quantum Ising model}

As a concrete example we now consider equilibration times and in particular
the prediction Eq.~(\ref{eq:T_eq_LE}) for the quantum Ising model
undergoing a sudden field quench. The Hamiltonian is
\begin{equation}
H=-\sum_{j=1}^{L}\left[\sigma_{j}^{x}\sigma_{j+1}^{x}+h\sigma_{j}^{z}\right],\label{eq:H_Ising}
\end{equation}
and periodic boundary conditions are used. The system is initialized
in the ground state of Eq.~(\ref{eq:H_Ising}) with parameter $h^{\left(1\right)}$.
At time $t=0$, the parameter is suddenly changed to $h^{\left(2\right)}$
and the state is let evolve unitarily with Hamiltonian $H\left(h_{2}\right)$.
The model in Eq.~(\ref{eq:H_Ising}) has critical points in the Ising
universality class at $h=\pm1$ with $d=\nu=\zeta=1$, separating
an ordered phase $\left|h\right|<1$ from a disordered paramagnetic
region $\left|h\right|>1$. For $\left|h\right|<1$ the order parameter
$\langle\sigma_{i}^{x}\rangle$ becomes non-zero, thus breaking the
$\mathbb{Z}_{2}$ symmetry ($\sigma_{i}^{x}\to-\sigma_{i}^{x}$) of
the Hamiltonian. 

According to Eq.~(\ref{eq:T_eq_LE}) and the discussion of the previous
section we expect the following behavior for the equilibration time
as a function of the quench parameters $h_{2},\,\delta h=h_{2}-h_{1}$
and system size $L$:
\begin{equation}
T_{\mathrm{eq}}\propto\begin{cases}
L & \mathrm{for}\,\, h_{2}=h_{c},\,\,\mathrm{and}\,\,\delta h\ll L^{-1}\\
\mathrm{const.} & \mathrm{otherwise}
\end{cases}\label{eq:Teq_Ising}
\end{equation}

\begin{figure}
\begin{centering}
\includegraphics[clip,width=8cm]{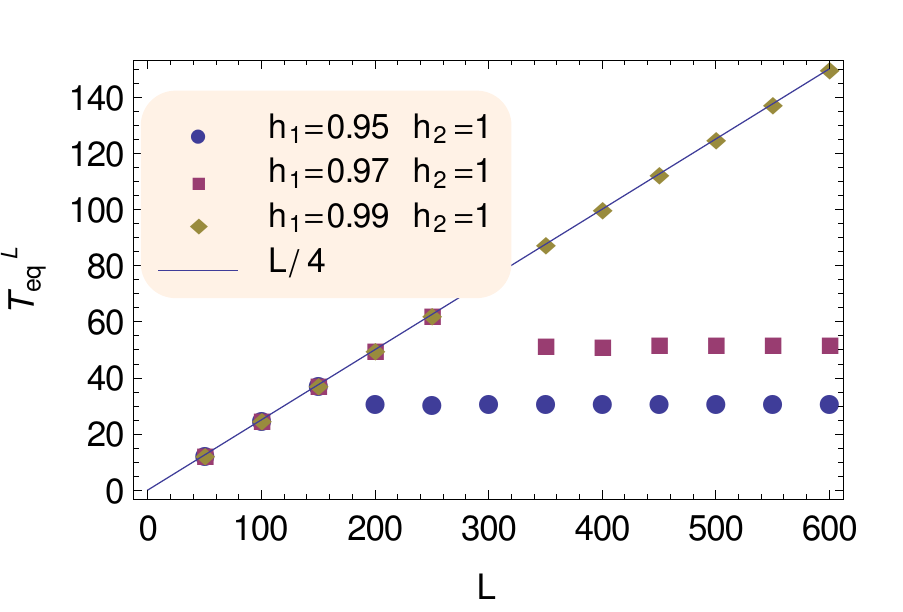}
\par\end{centering}

\caption{Equilibration times $T_{\mathrm{eq}}^{\mathcal{L}}$ for the Loschmidt
echo as a function of the system size $L$ for different quench parameters
$h_{1,2}$. $T_{\mathrm{eq}}^{\mathcal{L}}$ is computed by solving
numerically for the first solution of $\mathcal{L}\left(t\right)=\overline{\mathcal{L}}$.
\label{fig:Teq_LE}}
\end{figure}

\begin{figure}
\begin{centering}
\includegraphics[clip,width=8cm]{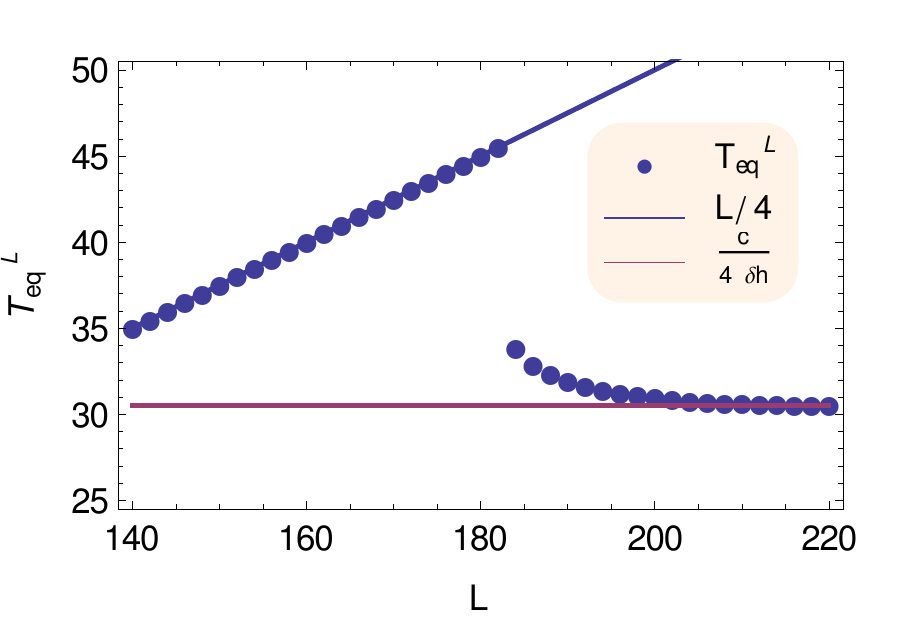}
\par\end{centering}

\caption{Equilibration times $T_{\mathrm{eq}}^{\mathcal{L}}$ for the Loschmidt
echo close to the crossover region $\delta h\simeq cL^{-1}$. \label{fig:Teq_LE_crossover}}
\end{figure}

As a first test we check whether Eq.~(\ref{eq:Teq_Ising}) is satisfied
for the Loschmidt echo itself. The LE for a sudden quench has been
computed in \cite{quan_decay_2006} (superscripts refer to to different
values of the coupling constants $h^{\left(i\right)}$)
\[
\mathcal{L}\left(t\right)=\prod_{k>0}\left[1-\sin^{2}\left(\delta\vartheta_{k}\right)\sin^{2}\left(\Lambda_{k}^{\left(2\right)}t/2\right)\right]
\]
where $\Lambda_{k}=2\sqrt{1+h^{2}+2h\cos(k)}$ is the single particle
dispersion, $\delta\vartheta_{k}=\vartheta_{k}^{\left(2\right)}-\vartheta_{k}^{\left(1\right)}$
and $\vartheta_{k}^{\left(i\right)}$ are the Bogoliubov angles at
parameters $h^{\left(i\right)}$, and the quasimomenta are quantized
according to $k=\pi\left(2n+1\right)/L,\, n=0,1,\ldots L/2-1$ (see
\cite{campos_venuti_unitary_2010} for further details). In Fig.~\ref{fig:Teq_LE}
we plot the equilibration time $T_{\mathrm{eq}}^{\mathcal{L}}$ of
the Loschmidt echo versus size for different quench parameters computed
exactly by solving numerically for the first solution of $\mathcal{L}\left(t\right)=\overline{\mathcal{L}}$.
Indeed Eq.~(\ref{eq:Teq_Ising}) is satisfied to a high accuracy.
Moreover the transition between the two behaviors of Eq.~(\ref{eq:Teq_Ising})
appears to be very sharp. From the numerical analysis the following
behavior for $T_{\mathrm{eq}}^{\mathcal{L}}$ emerges valid outside
the crossover region $\delta h\approx cL^{-1}$
\[
T_{\mathrm{eq}}^{\mathcal{L}}=\begin{cases}
\frac{L}{4} & \mathrm{for}\,\, h_{2}=h_{c},\,\,\mathrm{and}\,\,\delta h\ll cL^{-1}\\
\frac{c}{4\delta h} & \mathrm{otherwise}
\end{cases},
\]
where the constant $c$ in principle depends on $h_{1},\, h_{2}$,
but for small quenches close to the critical point tends to $c\simeq6.1$.
The typical behavior in the crossover region $\delta h\simeq cL^{-1}$
is depicted in Fig.~\ref{fig:Teq_LE_crossover}. 

It is natural to ask whether the predictions of Eq.~(\ref{eq:Teq_Ising})
are satisfied for observables other than the Loschmidt echo. To this
end, we consider the transverse magnetization $m^{z}\left(t\right)=\langle\sigma_{i}^{z}\left(t\right)\rangle$
which can be computed as \cite{barouch_statistical_1970,campos_venuti_unitary_2010}
\begin{multline}
m^{z}\left(t\right)=\frac{1}{L}\sum_{k}\cos\left(\vartheta_{k}^{\left(2\right)}\right)\cos\left(\delta\vartheta_{k}\right)\\
+\sin\left(\vartheta_{k}^{\left(2\right)}\right)\sin\left(\delta\vartheta_{k}\right)\cos\left(t\Lambda_{k}^{\left(2\right)}\right)\label{eq:mz}
\end{multline}

\begin{figure}
\begin{centering}
\includegraphics[clip,width=8cm]{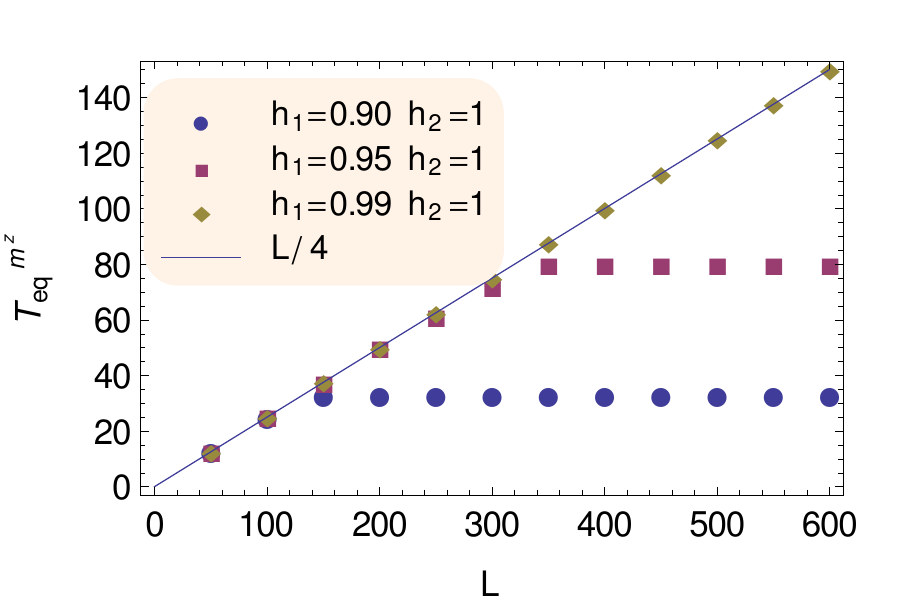}
\par\end{centering}

\caption{Equilibration times $T_{\mathrm{eq}}^{m^{z}}$ for the transverse
magnetization $m^{z}\left(t\right)=\langle\sigma_{1}^{z}(t)\rangle$
as a function of the size $L$ for different quench parameters $h_{1,2}$.
$T_{\mathrm{eq}}^{m^{z}}$ is computed exactly solving for the first
solution of $m^{z}\left(t\right)=\overline{m^{z}}$. \label{fig:Teq_mz}}
\end{figure}

From eq.~(\ref{eq:mz}) we extract the equilibration time from the
solution of $m^{z}\left(t\right)=\overline{m^{z}}=L^{-1}\sum_{k}\cos\left(\vartheta_{k}^{\left(2\right)}\right)\cos\left(\delta\vartheta_{k}\right)$.
The numerical results, shown in Fig.~\ref{fig:Teq_mz}, confirm that
Eq.~(\ref{eq:T_eq_LE}) is applicable to the transfer magnetization
too. The numerical results can be summarized as 
\[
T_{\mathrm{eq}}^{m^{z}}=\begin{cases}
\frac{L}{4} & \mathrm{for}\,\, h_{2}=h_{c},\,\,\mathrm{and}\,\,\delta h\ll c'L^{-1}\\
\frac{c'}{4\delta h} & \mathrm{otherwise}
\end{cases}
\]
with constant given now by $c'\simeq15$ for small quenches close
to criticality. 

\begin{figure}
\begin{centering}
\includegraphics[clip,width=8cm]{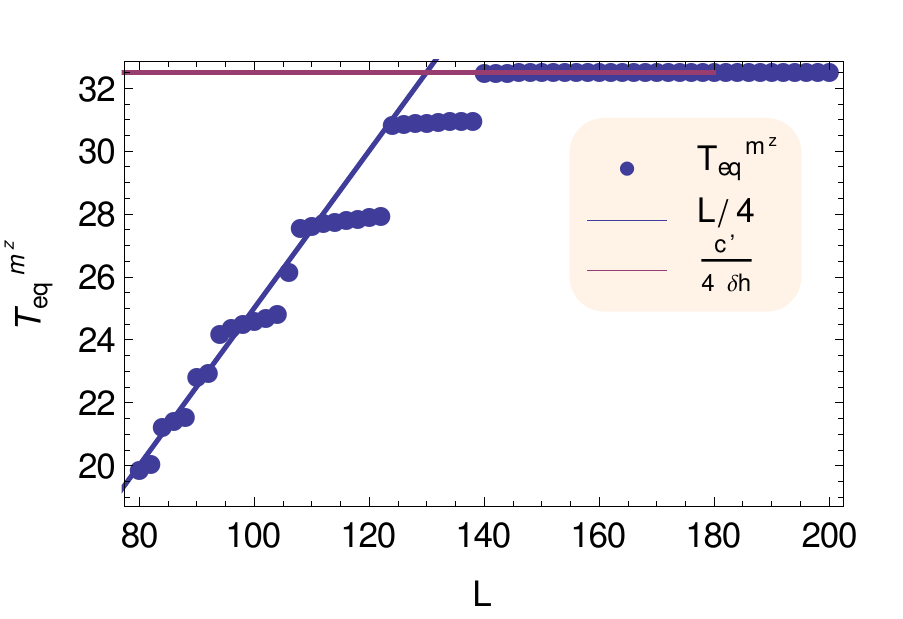}
\par\end{centering}

\caption{Equilibration times $T_{\mathrm{eq}}^{m^{z}}$ for the transverse
magnetization close to the crossover region $\delta h\simeq c'L^{-1}$.
For this values of parameters $c'\simeq13$. \label{fig:Teq_mz_crossover}}
\end{figure}

Finally, let us comment on the approach to equilibrium of the order
parameter $m^{x}\left(t\right)=\langle\sigma_{i}^{x}\left(t\right)\rangle$.
We recall that, since a non-zero $m^{x}$ breaks the symmetry of the
Hamiltonian, $m^{x}$ must be computed as the clustering part of an
equal time correlation function: $\langle\sigma_{i}^{x}\left(t\right)\sigma_{j}^{x}\left(t\right)\rangle\stackrel{\left|i-j\right|\to\infty}{\longrightarrow}\left[m^{x}\left(t\right)\right]^{2}$.
As usual in symmetry broken phases this requires the thermodynamic
limit to be taken first, but a finite size approximation for systems
with periodic boundary conditions can be obtained by considering the
correlation at half chain $\left[m_{L}^{x}\left(t\right)\right]^{2}:=\langle\sigma_{i}^{x}\left(t\right)\sigma_{i+L/2}^{x}\left(t\right)\rangle$. 

We do not expect formula (\ref{eq:T_eq_LE}) to reproduce the equilibration
time correctly for the order parameter, as it does not distinguish
whether we are in the ordered phase or not. 

The behavior of $m^{x}\left(t\right)$ has recently been obtained
analytically in the thermodynamic limit in the quench setting \cite{calabrese_quantum_2012}.
The results of \cite{calabrese_quantum_2012} for $m^{x}\left(t\right)$
can be summarized as follows. First of all $m^{x}\left(t\right)=0$
identically for quenches starting in the disordered phase $(\left|h_{1}\right|>1$)
as the symmetry remains unbroken. For quenches starting in the ordered
phase $(\left|h_{1}\right|<1)$ one has a different behavior depending
whether one ends up in the ordered or disordered phase. More specifically
\[
m^{x}\left(t\right)=\begin{cases}
Ae^{-t/\tau} & \left|h_{2}\right|<1\\
A'e^{-t/\tau}\sqrt{1+\cos\left(\Lambda_{k_{0}}t+\alpha\right)+\ldots} & \left|h_{2}\right|>1
\end{cases}
\]
The constants $A,\, A',\,\tau,\, k_{0},\,\alpha$ all depends on $h_{1},\, h_{2}$
and are given explicitly in \cite{calabrese_quantum_2012}. For quenches
starting end ending in the ordered region, the equation $m^{x}\left(t\right)=\overline{m^{x}}=m^{x}\left(\infty\right)=0$
has no real solution. Correspondingly at finite size, $T_{\mathrm{eq}}^{m^{x}}$
must be an increasing function of $L$ and the simplest guess is $T_{\mathrm{eq}}^{m^{x}}\propto L^{\zeta}=L$.
Instead for quenches ending in the disordered region, we see that
$m^{x}\left(t\right)=0$ has a finite solution even in the thermodynamic
limit, and so we expect in this case $T_{\mathrm{eq}}^{m^{x}}=O\left(L^{0}\right)$.

\section{The Anderson model}

We now turn to random systems. The model we consider is the tight-binding
model with random diagonal disorder, sometimes referred to as the
Anderson model \cite{anderson_absence_1958}:
\begin{equation}
H=\sum_{j=1}^{L}\left[t\left(c_{j}^{\dagger}c_{j+1}+c_{j+1}^{\dagger}c_{j}\right)-\mu_{i}c_{i}^{\dagger}c_{i}\right].\label{Ham}
\end{equation}
The diagonal elements $\mu_{i}$ are identically distributed independent
random variables. In all of our simulations we will use a uniform,
flat distribution in the interval $\left[-W,W\right]$. Through the
Jordan-Wigner mapping, Hamiltonian Eq.~(\ref{Ham}) equivalently
describes an $XX$ chain of $L$ spins in a random magnetic field.
The Hamiltonian Eq.~(\ref{Ham}) can be written in compact notation
as $H=c^{\dagger}M_{\mu}c$ , where $c=\left(c_{1},\ldots,c_{L}\right)^{T}$,
$M_{\mu}$ is the one particle Hamiltonian and the subscript $\mu$
refers to the random variables $\mu_{i}$. In the infinite volume
limit, the spectrum of $M_{\mu}$ is given, with probability one,
by $\sigma\left(\nabla^{2}\right)+\mathrm{supp}\left(-\mu_{i}\right)=\left[-2t,2t\right]+\mathrm{supp}\left(-\mu_{i}\right)$
where $\nabla^{2}$ is the discrete Laplacian in 1D, $\left(\nabla^{2}\right)_{i,j}=\delta_{i,j+1}+\delta_{i,j-1}$,
and $\mathrm{supp}$ is the support. In case of a  uniform distribution
we have $\sigma\left(M_{\mu}\right)=\left[-2t-W,2t+W\right]$. Moreover,
for any finite amount of randomness, the spectrum is almost surely
pure point, i.e.~consists only of eigenvalues, and the eigenfunctions
are exponentially localized (see e.g.~\cite{kunz_sur_1980,frohlich_constructive_1985}).
Such a situation is referred to as a \emph{localized} phase. Since
in the absence of disorder the model Eq.~(\ref{Ham}) is a band conductor,
in one spatial dimension there is a metal-insulator transition for
any however small amount of disorder $W>0$. 

To study the equilibration properties of the Anderson model, we proceed
as follows. We initialize the system in a state $\rho_{0}$, evolve
it unitarily with one instance of Hamiltonian Eq.~(\ref{Ham}) into
$\rho\left(t\right)$, and consider the expectation value of some
observable $A$: $\mathcal{A}\left(t\right)=\tr\left(A\rho\left(t\right)\right)$.
For very large systems one expects that concentration results will
apply and that the single instance $\mathcal{A}\left(t\right)$ will
be, with very large probability, close to its ensemble average $\mathsf{A}\left(t\right):=\mathsf{E}\left[\mathcal{A}\left(t\right)\right]$
where we denoted with $\mathsf{E}\left[\cdot\right]$ the average
over the random potentials $\mu_{i}$. When this happens, or more
precisely when the relative variance $\Delta^{2}\mathsf{A}/\mathsf{A}^{2}\to0$
as the system size increases, one says that $\mathcal{A}$ is self-averaging.
We will not be concerned with this issue here, we just notice that
$\mathsf{A}\left(t\right)$ generally gives the result of a hypothetical
measurement of $A$ within the confidence interval. The equilibration
time is then obtained by solving for the first solution of $\mathsf{A}(T_{\mathrm{eq}})=\overline{\mathsf{A}}$. 

To be concrete we will consider a Gaussian initial state $\rho_{0}$
with $N$ particles uniquely specified by the covariance matrix $R_{i,j}=\tr(c_{i}^{\dagger}c_{j}\rho_{0})$.
Since Eq.~(\ref{Ham}) conserves particle number, the evolution is
constrained to the sector with $N$ particles. As observables, we
choose a general quadratic operator given by $A=\sum_{i,j}a_{i,j}c_{i}^{\dagger}c_{j}$.
In this case the expectation value $\mathcal{A}\left(t\right)$ can
be completely characterized in terms of one-particle matrices \cite{campos_venuti_gaussian_2012}:
\begin{equation}
\mathcal{A}\left(t\right)=\tr\left[A\rho\left(t\right)\right]=\tr\left(ae^{-itM_{\mu}}R^{T}e^{itM_{\mu}}\right)\,.\label{eq:A_observable}
\end{equation}
To be specific we will study two particular quadratic observables:
$N_{\ell}:=\sum_{j=1}^{\ell}c_{i}^{\dagger}c_{i}$ which counts how
many particles are present in the first $\ell$ sites (say from left).
In this case the thermodynamic limit is given by fixing the particle
density $\nu=N/L$ together with the {}``observable'' density $\alpha=\ell/L$
and let $L\to\infty$. 

As previously argued, a useful quantity to consider is the LE. In
this free Fermionic setting it can be written as \cite{klich_full_2002,peschel_calculation_2003}
\[
\mathcal{L}\left(t\right)=\left|\det\left(1-R^{T}+R^{T}e^{-itM_{\mu}}\right)\right|^{2}\,.
\]

Because of the random nature of the problem we do not expect the initial
locations of the $N$ particles to matter particularly. Hence we will
consider an initial state where all the $N$ particles are pushed
to the left, i.e.~$R_{i,i}=1$ for $i=1,2,\ldots,N$ and all other
entries zero. We have checked that initializing the particles at other
sites does not change our results. In any case for pure initial states,
$R^{T}$ is a (orthogonal) projector $\left(R^{T}\right)^{2}=R^{T}$,
meaning its eigenvalues are either 1 or 0. Therefore, in some basis,
$R^{T}$ always has the aforementioned form, i.e.~there exists a
unitary matrix $X$ such that $X^{\dagger}R^{T}X=R_{N}^{T}=\mathrm{diag}\left(1,1,\ldots,1,0,\ldots,0\right)$
with $N$ entries $1$ and $\left(L-N\right)$ zeros. The Loschmidt
echo then becomes $\mathcal{L}\left(t\right)=\left|\det\left(1-R_{N}^{T}+R_{N}^{T}X^{\dagger}e^{-itM}X\right)\right|^{2}$.
Our choice of initial state corresponds to having $X=\1$. Thanks
to the simple form of $R_{N}^{T}$, one can evaluate the determinant
using Laplace's formula and reduce it to a determinant of an $N\times N$
matrix, i.e.
\[
\mathcal{L}\left(t\right)=\left|\det\Gamma_{N}\left[X^{\dagger}e^{-itM}X\right]\right|^{2}
\]
where the operator $\Gamma_{N}\left[Y\right]$ truncates the last
$L-N$ rows and columns of $Y$. Since clearly $\left\Vert \Gamma_{N}\left[U\right]\xi\right\Vert \le\left\Vert \xi\right\Vert $
for any unitary matrix $U$ and $L$-dimensional complex vector $\xi$,
the eigenvalues of $\Gamma_{N}\left[X^{\dagger}e^{-itM}X\right]$,
$z_{i}\left(t\right)$ have modulus smaller than one. The LE can than
be written as 
\[
\mathcal{L}\left(t\right)=e^{\mathcal{F}\left(t\right)}=e^{\nu Lf\left(t\right)}
\]
having defined $f\left(t\right)=\left(1/N\right)\sum_{i=1}^{N}\ln\left|z_{i}\left(t\right)\right|^{2}$.
Provided $f\left(t\right)$ has a limit as $L\to\infty$, this shows
that $\mathcal{L}\left(t\right)$ is exponentially small in the system
size. Accordingly, since $\mathcal{F}\left(t\right):=\ln\mathcal{L}\left(t\right)$
is extensive, we expect it to be self-averaging and convenient to
study. We also consider the averaged Loschimdt echo $\mathsf{L}\left(t\right)=\mathsf{E}\left[\mathcal{L}\left(t\right)\right]$
and the ensemble average of the logarithm of the Loschimidt echo (LLE)
$\mathsf{F}\left(t\right)=\mathsf{E}\left[\mathcal{F}\left(t\right)\right]$.
Note that by Jensen's inequality $\mathsf{L}\left(t\right)\ge e^{\mathsf{F}\left(t\right)}$. 

\begin{figure}
\begin{centering}
\includegraphics[clip,scale=0.2]{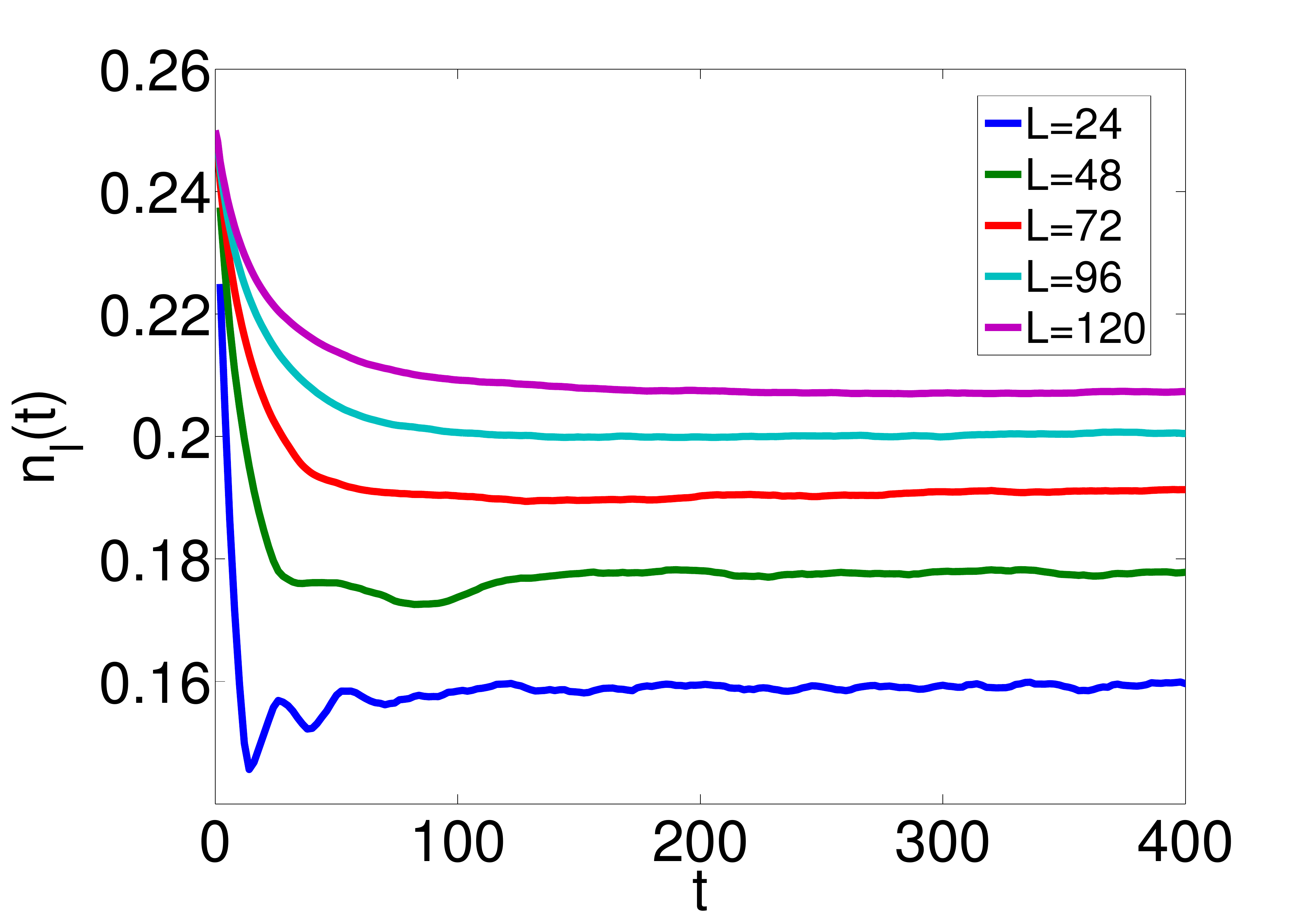}
\par\end{centering}

\caption{Relaxation of $\mathsf{n}_{\ell}\left(t\right)=\mathsf{E}\left[N_{\ell}\left(t\right)/L\right]$
with time for lengths $L=16,\,40,\,64,\,112$, at filling $\nu=N/L=0.25$
and $\alpha=\ell/L=0.5$. The random distribution has variance $\sigma=0.6$
($W=0.6\sqrt{3}\approx1.03$). The ensemble average is computed using
1000 realizations at each time.\label{fig:A_vs_time}}
\end{figure}

In Fig.~\ref{fig:A_vs_time} we show the results of our numerical
simulations for the --ensemble averaged-- observable $\mathsf{n}_{\ell}\left(t\right)=\mathsf{E}\left[N_{\ell}\left(t\right)/L\right]$.
When computing the equilibration time for observable $A$ by looking
for the first solution of $\mathsf{A}(T_{\mathrm{eq}})=\overline{\mathsf{A}}$,
the computationally most demanding part is the calculation of $\overline{\mathsf{A}}$,
especially for large system sizes. For quadratic observables, this
computation can be simplified considerably. We first note that the
time and ensemble averages commute. This is essentially a consequence
of Fubini's theorem and the fact that all our quantities are bounded.
The time average of $\mathcal{A}\left(t\right)$ for a particular
realization can be computed diagonalizing $M_{\mu}$. With the notation
$M_{\mu}=Q^{\dagger}\Lambda Q$, $\Lambda=\mathrm{diag}\left\{ \epsilon_{1},\ldots,\epsilon_{L}\right\} $,
and $Q$ unitary, one has $\mathcal{A}\left(t\right)=\sum_{n,m}\left(Q^{\dagger}AQ\right)_{n,m}\left(Q^{\dagger}R^{T}Q\right)_{m,n}e^{-it\left(\epsilon_{m}-\epsilon_{n}\right)}$.
Now, simply note that, with probability one, the spectrum $\epsilon_{n}$
is non-degenerate. This implies that the time average is given exactly
by
\begin{equation}
\overline{\mathcal{A}}=\sum_{n}\left(Q^{\dagger}AQ\right)_{n,n}\left(Q^{\dagger}R^{T}Q\right)_{n,n}\label{eq:A_average}
\end{equation}

\begin{figure}
\begin{centering}
\includegraphics[scale=0.2]{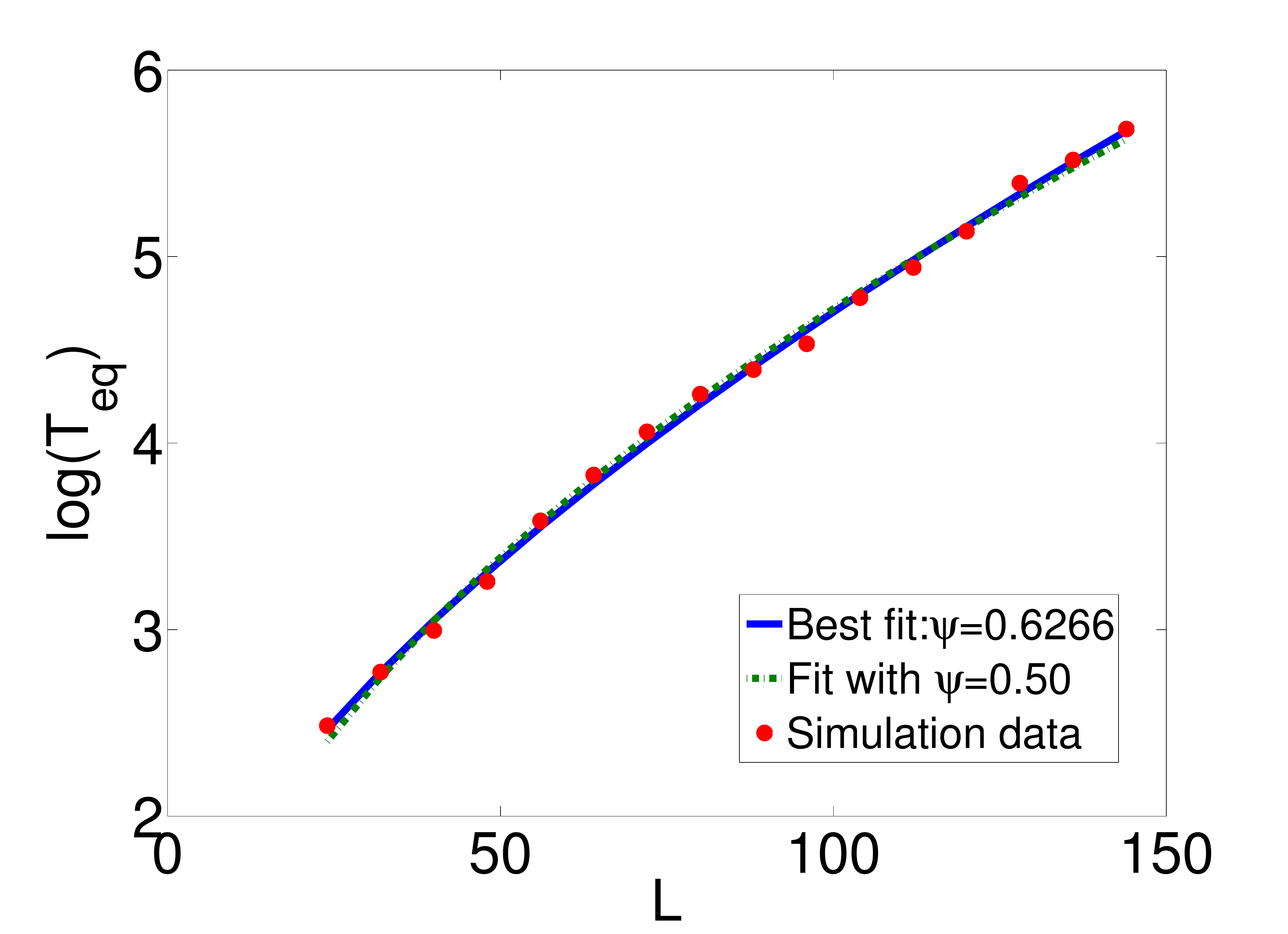}
\par\end{centering}

\caption{Scaling of equilibration time $T_{\mathrm{eq}}(L)$ vs system size
$L$, extracted from $\mathsf{n}_{\ell}\left(t\right)$. $T_{\mathrm{eq}}$
is obtained from this first solution of $\mathsf{E}[n_{\ell}(T_{\mathrm{eq}})]=\overline{\mathsf{E}[n_{\ell}(t)]}$.
Parameters are as in Figure \ref{fig:A_vs_time} . The ensemble averages
are obtained summing over 2000 realizations. The equilibration time
is found to obey $\log(T_{\mathrm{eq}})=0.021L^{0.6266}+2.53$ with
residual norm $0.17$.\label{fig:Tl_scaling_for_nl}}
\end{figure}
The value $\overline{\mathsf{A}}$ is then computed taking the ensemble
average of $\overline{\mathcal{A}}$, $\overline{\mathsf{A}}=\mathsf{E}\left[\overline{\mathcal{A}}\right]$,
using Eq.~(\ref{eq:A_average}). In figure \ref{fig:Tl_scaling_for_nl}
we plot the equilibration time obtained for $\mathsf{n}_{\ell}\left(t\right)$
as a function of system size $L$. Our numerical results show that
the equilibration time scales exponentially in the system size. 

For the case of the LE and the LLE we have not been able to compute
the time average exactly. It is known that in case of non-degenerate
many body energies, which is safe to assume in presence of randomness,
one has $\overline{\mathcal{L}}=\sum_{n}\left|\langle\psi_{0}|\left\{ n\right\} \rangle\right|^{4}$
\cite{campos_venuti_unitary_2010} (where $|\left\{ n\right\} \rangle=|n_{1},\ldots,n_{L}\rangle$'s
are the many-body eigenfunctions satisfying $\sum_{j}n_{j}=N$). For
our choice of initial state, the amplitudes are given by
\[
\langle\psi_{0}|\left\{ n\right\} \rangle=\det V_{\left[1,2,\ldots,N\right]}Q^{\dagger}V_{\left\{ n\right\} }^{\dagger}
\]
where $V_{\left[1,\ldots,N\right]}$ is the $N\times L$ matrix with
ones on the diagonal and zero otherwise. $V_{\left\{ n\right\} }$
is formed in the same way but the ones on the diagonal are in correspondence
of the row $i$ for which $n_{i}=1$. The normalization of the weights
is provided by Cauchy-Binet's formula
\begin{multline*}
\sum_{\stackrel{\left\{ n\right\} }{\sum_{i}n_{i}=N}}\det V_{\left[1,2,\ldots,N\right]}Q^{\dagger}V_{\left\{ n\right\} }^{\dagger}\det V_{\left\{ n\right\} }TQV\\
=\det V_{\left[1,2,\ldots,N\right]}Q^{\dagger}QV_{\left[1,2,\ldots,N\right]}^{\dagger}=1
\end{multline*}
The time average Loschmidt echo is then given by
\[
\overline{\mathcal{L}}=\sum_{\stackrel{\left\{ n\right\} }{\sum_{i}n_{i}=N}}\left|\det V_{\left[1,2,\ldots,N\right]}Q^{\dagger}V_{\left\{ n\right\} }^{\dagger}\right|^{4}
\]
The above sum, however, contains an exponential number of terms and
is not practical for numerical computations. To evaluate $\overline{\mathsf{L}}$
and $\overline{\mathsf{F}}$, we compute the $\mu$-random average
$\mathsf{L}\left(t\right)\simeq N_{\mathrm{samples}}^{-1}\sum_{i=1}^{N_{\mathrm{samples}}}\mathcal{L}_{i}\left(t\right)$
(and similarly for $\mathsf{F}$) using as many as $N_{\mathrm{samples}}=2000$
for $N_{\mathrm{times}}$ random times uniformly distributed between
$\left[T_{0},T\right]$. The time average is then obtained via $\overline{\mathsf{L}}\simeq N_{\mathrm{samples}}^{-1}N_{\mathrm{times}}^{-1}\left(T-T_{0}\right)^{-1}\sum_{j=1}^{N_{\mathrm{times}}}\sum_{i=1}^{N_{\mathrm{samples}}}\mathcal{L}_{i}(t_{j})$.
We use a positive $T_{0}$ to get rid of the initial transient and
obtain more precise estimates with the same computational cost. Typical
valued we used are $T\simeq1000$, $T_{0}\simeq750$. \textcolor{black}{Larger
times of the order of }$T\simeq10000$\textcolor{black}{{} with }$T_{0}\simeq1000$\textcolor{black}{{}
were used for systems of size exceeding }$L\sim128$\textcolor{black}{.}
\begin{figure}
\begin{centering}
\includegraphics[clip,scale=0.2]{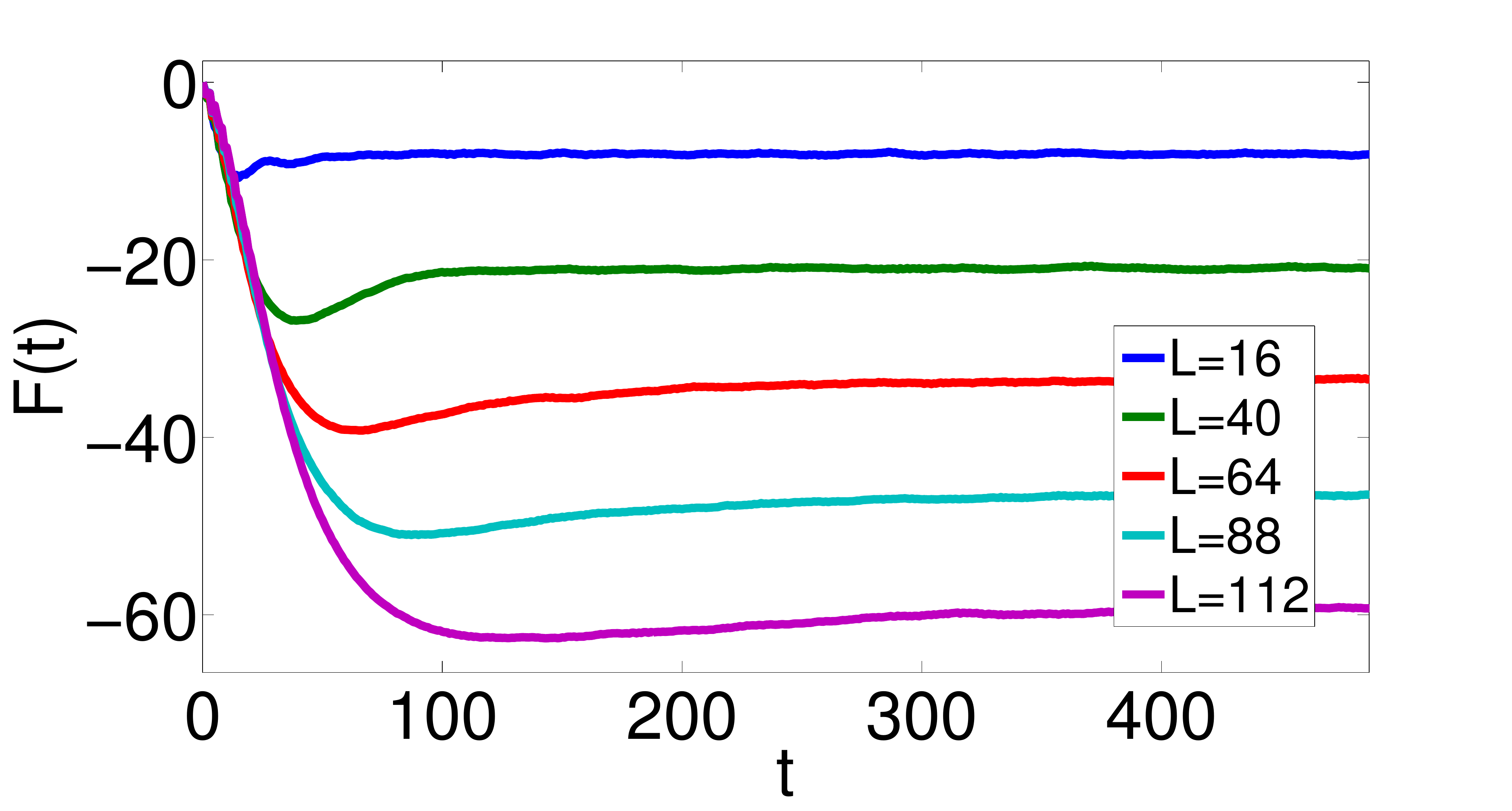}
\par\end{centering}

\caption{Relaxation of $\mathsf{F}(t)=\mathsf{E}[\log(\mathcal{L}(t))]$. Parameters
are as in Figure \ref{fig:A_vs_time} and an ensemble average was
performed over 1000 realizations at each instant in time. \label{fig:loglos_vs_time}}
\end{figure}

\begin{figure}
\begin{centering}
\includegraphics[scale=0.2]{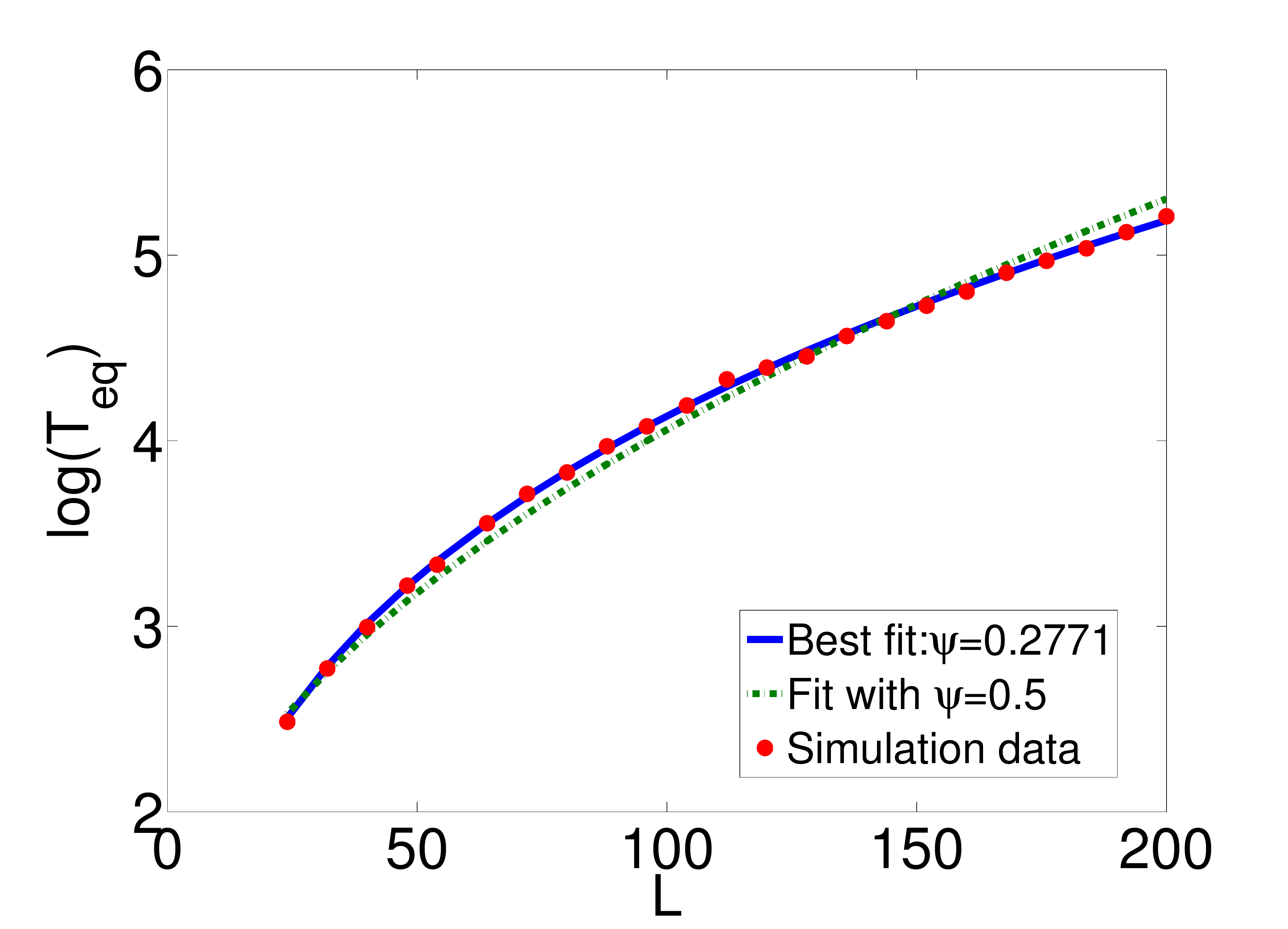}
\par\end{centering}

\caption{Scaling of the relaxation time $T_{eq}$ such that $\mathsf{F}(T_{eq})=\overline{\mathsf{F}}$
\label{fig:Tl_vs_L_for_loglos}, $T_{eq}=1.39L^{0.2771}-0.847$ with
residual norm $0.05$. }
\end{figure}

In figure \ref{fig:loglos_vs_time} we plot the ensemble averaged
LLE time series $\mathsf{F}\left(t\right)$ at different sizes $L$
while in figure \ref{fig:Tl_vs_L_for_loglos} we show the equilibration
times of the logarithmic Loschmidt echo  as a function of $L$ .
In this case as well, the numerical results indicate equilibration
times diverging exponentially in the system size. Based on the numerical
evidence, we conjecture that for more general observables $O$, the
equilibration time in the Anderson model might satisfy 
\[
\ln T_{\mathrm{eq}}^{O}=c_{O}L^{\psi}+d_{O}
\]
 where $c_{O},\, d_{O},\,\psi$ are constants which depend on the
observable $O$. 

\section{A CUE model}

\begin{figure}
\begin{centering}
\includegraphics[width=8cm,height=5cm]{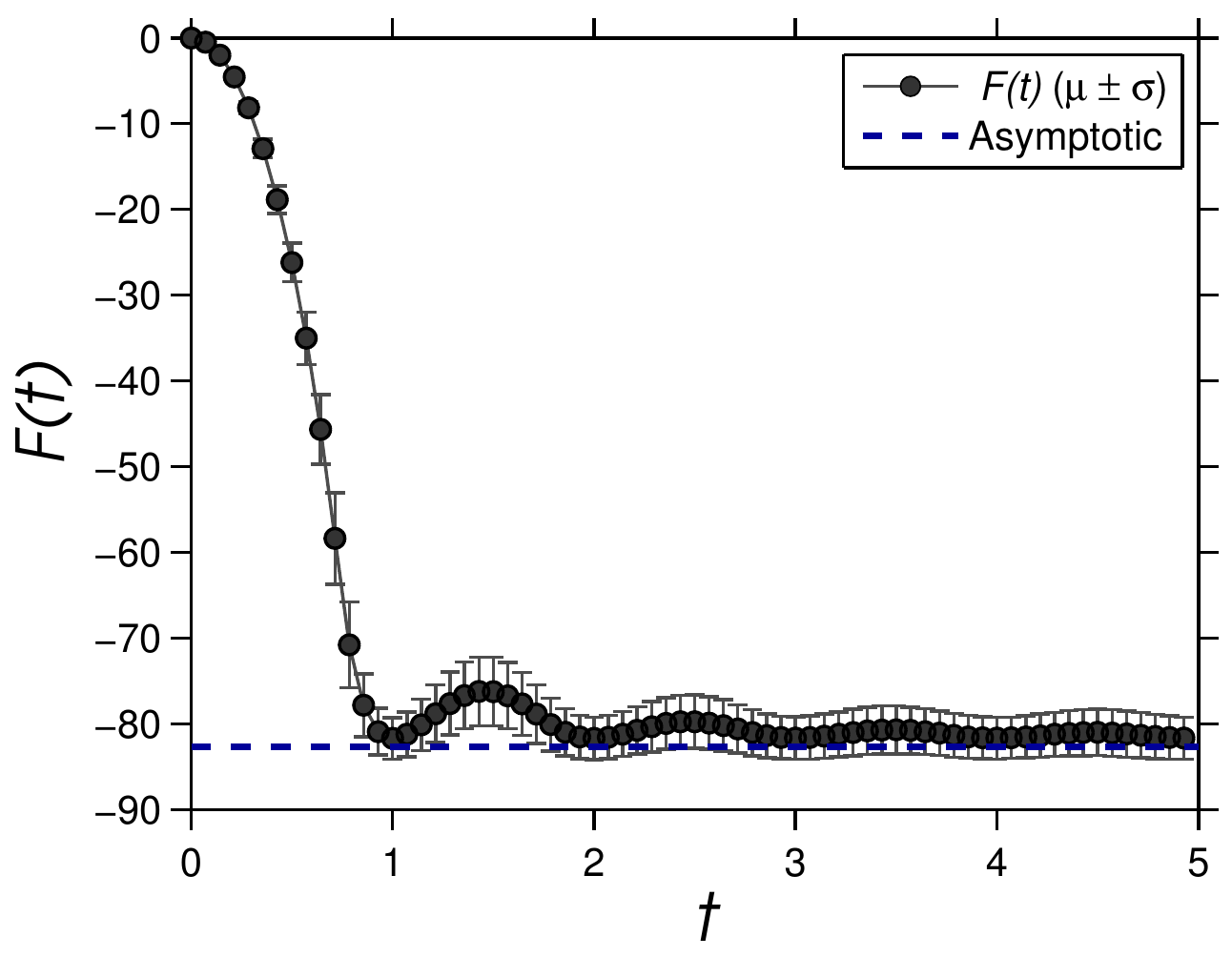}
\par\end{centering}

\caption{Plot of $\mathsf{F}\left(t\right)$ for $L=120$, at $\nu=1/2$ (half
filling). The variance $\delta\mathsf{F}\left(t\right)$ is shown
as error bar. Numerical results are obtained averaging over $N_{\mathrm{samples}}=7500$
unitaries distributed according to the Haar measure and obtained with
the algorithm outlined in \cite{mezzadri_how_2006}.\label{fig:Plot-of-F}}
\end{figure}

In this section we consider another random matrix model for which
we are able to prove analytically that $T_{\mathrm{eq}}\le1$. In
this model we fix the evolution operator $U$ at time $t=1$ to be
taken from the circular unitary ensemble (CUE). This means that $U$
is an $L\times L$ unitary matrix sampled from the uniform Haar measure
over the group $\mathbb{U}\left(L\right)$. At other times the evolution
is defined via $U^{t}$. The arbitrariness in the definition of $U^{t}$
for $t\in\mathbb{R}$ is fixed in the following way. Any unitary matrix
$U$ from the CUE can be written as $U=V^{\dagger}e^{i\boldsymbol{\phi}}V$
where $V$ is again Haar distributed, $e^{i\boldsymbol{\phi}}:=\mathrm{diag}(e^{i\phi_{1}},e^{i\phi_{2}},\ldots,e^{i\phi_{L}})$
and the phases $\phi_{i}$ are distributed according to $P\left(\boldsymbol{\phi}\right)=C\prod_{i<j}\left|e^{i\phi_{i}}-e^{i\phi_{j}}\right|^{2}$
where $C^{-1}=\left(2\pi\right)^{L}L!$ is the normalization constant
and $\phi_{i}\in[0,2\pi)$ (see e.g.~\cite{mehta_random_2004}).
Our model is described by taking a Hamiltonian $H=\boldsymbol{\phi}:=\mathrm{diag}(\phi_{1},\phi_{2},\ldots,\phi_{L})$
and considering the average over all isospectral Hamiltonians $H'=V^{\dagger}HV$
with $V$ Haar distributed. The dynamical evolution is given by $e^{itH'}$.
Given these considerations this model is equivalent to the ensemble
considered in \cite{brandao_convergence_2011,cramer_thermalization_2012}
after averaging all the energies $E_{i}$ with the CUE distribution
$P\left(\boldsymbol{\phi}\right)$. Let us now turn to the computation
of $\mathsf{F}\left(t\right)=\mathsf{E}_{U}\left[\ln\det\Gamma_{N}\left(e^{itH'}\right)\right]$.
If $t=n$ is an integer, the evolution operator is given by $e^{itH'}=U^{n}$
where $U^{n}$ is unitary and Haar distributed, hence, at integer
times we obtain 
\[
\mathsf{F}\left(n\right)\equiv\mathsf{F}_{0}=\mathsf{E}_{U}\left[\ln\left|\det\Gamma_{N}(U)\right|^{2}\right],
\]
independent of $n$. We observed, as it is natural to expect, that
the function $\mathsf{F}\left(t\right)$ has a limit as $t\to\infty$
(see Fig.~\ref{fig:Plot-of-F}). In this case the limit must coincide
with the time average and with $\mathsf{F}(n)$ for $n$ integer,
i.e.~$\mathsf{F}\left(\infty\right)=\mathsf{F}(n)=\overline{\mathsf{F}}$.
This shows that the relaxation time in this random systems is bounded
by one. In fact in principle one could have $\mathsf{F}\left(t\right)=\overline{\mathsf{F}}$
also for a time $t$ smaller than one, however our numerics indicates
that the first occurrence of $\mathsf{F}\left(t\right)=\overline{\mathsf{F}}$
is indeed at $T_{\mathrm{eq}}=1$ (see Fig.~\ref{fig:Plot-of-F})
independent of the system size. We would like to mention at this point
the results of Ref.~\cite{brandino_quench_2012} where a different
sparse random ensemble has been constructed and an equilibration time
growing as the system size has been reported. These findings show
that random systems can give rise, in general, both to slow and fast
equilibration processes and the correct equilibration time-scale can
only be obtained through an accurate investigation (although one expects
faster equilibration to be associated to less sparse Hamiltonians). 

It turns out that the limiting value $\mathsf{F}\left(\infty\right)$
can be obtained exactly. The distribution of eigenvalues of truncated
matrices $\Gamma_{N}(U)$ when $U$ is Haar distributed has been computed
in \cite{zyczkowski_truncations_2000}. The eigenvalues $z_{i}$ of
$\Gamma_{N}(U)$ are complex numbers in the unit disk $\left|z_{i}\right|\le1$.
Calling $r_{i}=\left|z_{i}\right|$, and defining the probability
distribution of the moduli $P_{L,N}\left(r\right)\equiv N^{-1}\sum_{i=1}^{N}\mathsf{E}_{U}\left[\delta\left(r-\left|z_{i}\right|\right)\right]$,
the CUE average at integers time is then given by
\begin{equation}
\mathsf{F}_{0}=N\int_{0}^{1}P_{L,N}\left(r\right)\ln\left(r^{2}\right)dr.\label{eq:F_ave}
\end{equation}
Zyczkowski and Sommers were able to compute the distribution of the
moduli $P_{L,N}(r)$ and obtained, with $x=r^{2}$
\[
P\left(r\right)=\frac{2r}{N}\frac{\left(1-x\right)^{L-N-1}}{\left(L-N-1\right)!}\left(\frac{d}{dx}\right)^{L-N}\frac{1-x^{L}}{1-x}.
\]
The probability density $P_{\nu}\left(r\right)$ in the thermodynamic
limit at fixed particle density $N/L=\nu,\, L\to\infty$, was also
computed in \cite{zyczkowski_truncations_2000} and is given by
\begin{equation}
P_{\nu}\left(r\right)=\left(\nu^{-1}-1\right)\frac{2r}{\left(1-r^{2}\right)^{2}}\label{eq:P_nu}
\end{equation}
for $r<\sqrt{\nu}$ and zero otherwise. Plugging Eq.~(\ref{eq:P_nu})
into Eq.~(\ref{eq:F_ave}) one obtains, in the thermodynamic limit,
the following particularly simple result
\begin{equation}
\mathsf{F}\left(\infty\right)=L\left[\nu\ln\nu+\left(1-\nu\right)\ln\left(1-\nu\right)\right]\,.\label{eq:F_entropy}
\end{equation}

Quite surprisingly Eq.~(\ref{eq:F_entropy}) is the negative Von
Neumann entropy of the Gaussian state with covariance matrix $\mathsf{R}\left(n\right)=\mathsf{E}\left[R\left(n\right)\right]$
obtained taking the ensemble average of the covariance matrix at integer
times $t=n$. The reasoning is the following. First we remind that
the von Neumann entropy of a Fermionic Gaussian state $\rho_{W}$
with covariance matrix $W$ is given by
\begin{multline*}
-S_{\mathrm{vN}}\left(\rho_{W}\right)=\tr\rho_{W}\ln\rho_{W}=\\
\tr W\ln W+\tr\left(\1-W\right)\ln\left(\1-W\right).
\end{multline*}
Now note that at integer times the evolution operator is $U^{n}=U'$
with $U'$ again unitary. At such times the average of the covariance
is then $\mathsf{R}\left(n\right)=\mathsf{E}_{U}\left[URU^{\dagger}\right]$
and is proportional to the identity by Schur's lemma. The constant
is fixed noting that $\tr\mathsf{R}\left(n\right)=\tr R=N$, which
implies $\mathsf{R}\left(n\right)=\nu\1$. The claim then follows
trivially taking traces of diagonal operators of the form $(\gamma\1)\ln(\gamma\1)$.
We do not know the average covariance at non-integer times, however
if a limit $t\to\infty$ exist it must coincide with $\mathsf{R}\left(n\right)$.
In that case we would have $\mathsf{F}\left(\infty\right)=-S\left(\rho_{\mathsf{R}(\infty)}\right)\,[=-S\left(\rho_{\mathsf{R}(n)}\right)]$.
Research is in progress to check weather this connection between the
average of the logarithmic Loschmidt echo and the von Newmann entropy
has more general validity.

Finally, we also verified that $\mathsf{F}\left(t\right)$ is indeed
self averaging, i.e.~the relative variance goes to zero as $L$ increases.
In particular, since $\mathsf{F}\left(t\right)\propto L,$ the variance
of the rescaled variable $\mathsf{F}\left(t\right)/L$ goes to zero.

\section{Conclusions}

In this paper we considered equilibration in finite-dimensional isolated
systems, and particularly concentrated on the time needed to reach
equilibrium and its scaling behavior with the system size. The standard
definition of equilibration time extracted from the exponential decay
of some observable does not work in finite system because the dynamics
is quasi-periodic and thus, no exponential decay can take place. Our
definition of equilibration time $T_{\mathrm{eq}}$ is precisely the
time needed for an observable $\mathcal{A}\left(t\right)=\langle\psi_{0}|A\left(t\right)|\psi_{0}\rangle$
to reach its equilibrium value, and is given by the earliest solution
of $\mathcal{A}\left(t\right)=\overline{\mathcal{A}}$. We first examined
clean systems. Considering the Loschmidt echo as a particular observable,
we showed that in the general situation of a gapped or clustering
initial state, the equilibration time is independent of system size,
i.e.~$T_{\mathrm{eq}}=O\left(L^{0}\right)$. On the other hand, for
small quenches close to a critical point, one finds $T_{\mathrm{eq}}=O\left(L^{\zeta}\right)$
where $\zeta$ is the dynamical exponent. We then turned to random
systems and tackled the tight-binding model with diagonal impurities
as an example. Considering different observables $A$, we found in
all cases, that $\ln T_{\mathrm{eq}}=c_{A}L^{\psi_{A}}+d_{A}$ , where
$c_{A},d_{A},\psi_{A}$ are observable-dependent constants. The exponential
divergence of equilibration time, however, seems a general feature
of the localized phase in this system. Finally we introduced a novel
random matrix model (similar to the one considered in \cite{brandao_convergence_2011,cramer_thermalization_2012})
for which we were able to prove that $T_{\mathrm{eq}}=O\left(1\right)$.
This shows that, obviously, a higher degree of randomization can help
the system reach equilibrium faster.

LCV wishes to thank Marcos Rigol for bringing to his attention ref.~\cite{gramsch_quenches_2012}
prior to publication and Vladimir Yurovsky for interesting discussions.
This work was partially supported by the ARO MURI grant W911NF-11-1-0268,
and by the National Science Foundation under Grant No. NSF PHY11-25915.

\bibliographystyle{unsrt}
\bibliography{equilibration_random}

\end{document}